\pdfoutput=1
\documentclass[journal]{IEEEtran}
\usepackage{dblfloatfix}

\usepackage{cite}
\usepackage[cmex10]{amsmath}
\usepackage[pdftex]{graphicx}
\graphicspath{{figs/}}
\DeclareGraphicsExtensions{.pdf}
\usepackage{array}
\usepackage{mathtools}
\usepackage{amssymb}
\usepackage{nicefrac}
\usepackage{enumitem}
\usepackage{booktabs}
\usepackage{url}
\usepackage{trfsigns}
\usepackage{trsym}
\usepackage{mathptmx}
\usepackage{subfigure}

\newcommand{\textopt}{\mathrm{o}}

\newcommand{\rbracks}[1]{{\left(#1\right)}}
\newcommand{\sbracks}[1]{{\left[#1\right]}}
\newcommand{\cbracks}[1]{{\left\{#1\right\}}}
\newcommand{\T}{\mathsf{T}}
\newcommand{\CT}{\mathsf{H}}
\newcommand{\ie}{i.e.}
\newcommand{\eg}{e.g.}
\newcommand{\R}{\mathbb{R}}

\newcommand{\norm}[1]{{\left\lVert#1\right\rVert}}
\newcommand{\E}{\operatorname{E}}
\newcommand{\abs}[1]{{\left\lvert#1\right\rvert}}
\newcommand{\ERBS}{\operatorname{ERBS}}
\newcommand{\etc}{etc.}
\newcommand{\RMSD}{\operatorname{RMSD}}
\newcommand{\SNRF}{\operatorname{SNRF}}

\newcommand{\wrt}{w.r.t.}

\renewcommand{\vec}[1]{\mathbf{#1}}

\begin{document}
%
\title{SISO and SIMO Accompaniment Cancellation \\for Live Solo Recordings Based on \\Short-Time ERB-Band Wiener Filtering \\and Spectral Subtraction}
%
%
%

\author{Stanislaw~Gorlow,~\IEEEmembership{Member,~IEEE,}
	Mathieu~Ramona, 
	and~Fran\c{c}ois~Pachet
\thanks{This work was funded by the European Commission under program FP7-ICT-2011-8 within the scope of the PRAISE project (EU FP7 318770).}%
\thanks{S. Gorlow is with Sony Computer Science Laboratory (CSL), 6 rue Amyot, 75005 Paris, \^Ile-de-France, France e-mail: gorlow@csl.sony.fr.}
\thanks{M. Ramona and F. Pachet are with Sony CSL Paris.}
\thanks{Manuscript originally submitted June 26, 2015.}%
}

\maketitle

\begin{abstract}
Research in collaborative music learning is subject to unresolved problems demanding new technological solutions. One such problem poses the suppression of the accompaniment in a live recording of a performance during practice, which can be for the purposes of self-assessment or further machine-aided analysis. Being able to separate a solo from the accompaniment allows to create learning agents that may act as personal tutors and help the apprentice improve his or her technique. First, we start from the classical adaptive noise cancelling approach, and adjust it to the problem at hand. In a second step, we compare some adaptive and Wiener filtering approaches and assess their performances on the task. Our findings underpin that adaptive filtering is inapt of dealing with music signals and that Wiener filtering in the short-time Fourier transform domain is a much more effective approach. In addition, it is very cheap if carried out in the frequency bands of auditory filters. A double-output extension based on maximal-ratio combining is also proposed.
\end{abstract}


%
\IEEEpeerreviewmaketitle

\section{Introduction}
\label{sec:intro}

\IEEEPARstart{P}{racticing} a musical instrument is usually associated with professional supervision and personalized feedback when it comes to an unskilled apprentice. This is particularly true for a novice. Otherwise, fatigue may set in quickly, and even the most talented student can lose interest in continuing with practice or even in learning music as such. But yet, not everybody is willing  to pay a personal tutor, especially if the outcome is unclear. Other factors, such as dispensability, can also influence one's decision. A reasonable compromise may consist in learning agents that take the role of the tutor. And to avoid further spendings on expensive hardware, the agents would preferably be installed on a tablet computer, which as of today is equipped with a speaker and a microphone.

Practicing, \eg, the jazz guitar, one of the main obstacles one would surely encounter from a signal processing point of view is the isolation of the solo from the recording, which as a rule contains the solo and the accompaniment. The latter is generated by the computer and can be deemed known. Thus, the challenging nature of the task stems from the fact that the accompaniment signal is altered by the speaker, the acoustic channel, and the microphone. Furthermore, the high spectral dynamics of musical signals and their high bandwidth render the task problematic for classic solutions. Adaptive filtering, \eg, is to be questioned whether it is capable of keeping pace with the music signal dynamics. Wiener filtering, then again, is probably very expensive given the high system order. The question is whether there is a compromise solution that could allow to bypass the pragmatic headphones solution, which is hostile to self-assessment and self-adjustment during play.

Prior to elaborating the accompaniment cancellation task, we revisit the main principles of adaptive noise cancellation, which is pertinent to our problem. We develop as a reference the corresponding asymptotically optimal Wiener filter in the time domain. Beyond that, it is shown how the subtraction of the accompaniment can be carried out in the frequency or the spectral domain. Then, we devote ourselves to the main task and propose an efficient solution, which is based on a simple transmission model. Subsequently, we extend our solution to the imaginary case where we possess custom hardware in the form of a tiny two-element microphone array, and propose a method to combine its outputs. The different approaches and their derivatives are simulated on a jazz guitar solo recording and compared with each other against four metrics. Beyond, the proposed algorithm is thoroughly evaluated and discussed in a separate task. We conclude the paper by showing up our major findings. Other issues such as noise reduction, \etc, are not considered in this work.

\section{Adaptive Noise Cancellation}
\label{sec:anc}

Adaptive noise cancellation, or cancelling, is a known signal processing technique widely used to suppress additive noise in a  corrupted signal \cite{Widrow1975, Widrow1985}. It requires a reference noise input that is strongly correlated with the corrupting noise signal to work properly. In this section, we revisit the main principles of the underlying idea before establishing a connection to our actual problem, which is accompaniment cancellation. 

\subsection{Signal Model}

\begin{figure}[!ht]
\centering
\includegraphics[width=\columnwidth]{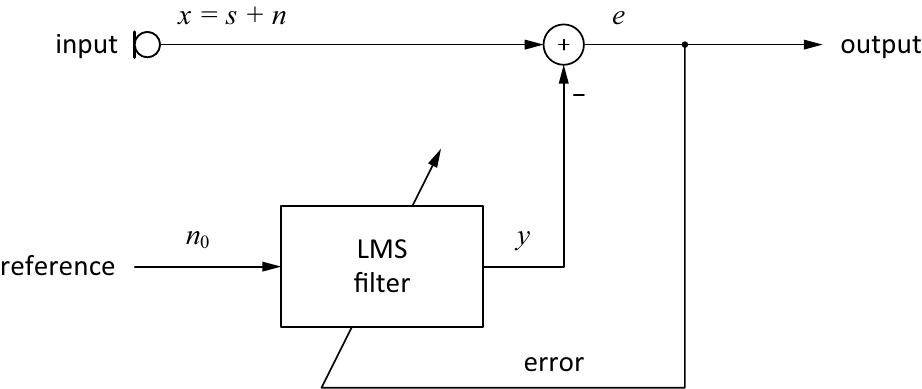}
\caption{Adaptive noise cancellation as a block diagram.}
\label{fig:anc}
\end{figure}

Fig.~\ref{fig:anc} depicts the operation of an adaptive noise canceller in the form of a block diagram. The reference noise $n_0$ is fed to the adaptive least mean squares (LMS) filter, which produces the estimate $y$ and subtracts it from the noisy input $s + n$. As it was mentioned before, $n_0$ and $n$ are correlated, whereas $s$ and $n$ are uncorrelated. All variables are Gaussian with zero mean. From the filter's point of view, the noisy input $x$ is the desired response while the output acts as the error that drives the adaptation. By minimizing the error $e$ in the least-squares sense, the adaptive filter provides the best fit of $n_0$ for $n$ in $x$. At the same time, the error, or output, $e$ converges to the best least-squares estimate for the signal $s$.

\subsection{Adaptive LMS Filtering}
\label{sec:lms}

The LMS filter is given by the  following recurrence relation:
\begin{equation}
\begin{aligned}
\vec{w}(k + 1) &= \vec{w}(k) + \mu \, \vec{n}_0(k) \, e(k) \qquad \text{with} \\ 
e(k) &= x(k) - \underbrace{\vec{w}^\T(k) \, \vec{n}_0(k)}_{y(k)} \text{,} 
\end{aligned}
\label{eq:lms}
\end{equation}
where $\vec{w}(k) \in \R^M$ is the weight vector at time instant $k$, and $\vec{w}(k +1)$ is the anticipated weight vector, respectively, $e(k)$ is the adaptation error, \ie, the difference between the observed signal $x(k)$ and the noise estimate $y(k)$, $\vec{n}_0(k) \in \R^M$ is thus the input noise sequence, and $\mu$ is a design parameter, which determines stability, rate of convergence, \etc\ \cite{Widrow1985}. The scale-invariant version of the LMS filter, which is insensitive to the scaling of the input, $n_0(k)$, is obtained by replacing 
\begin{equation}
\vec{n}_0(k) \leftarrow \frac{\vec{n}_0(k)}{\norm{\vec{n}_0(k)}_2^2} 
\end{equation}
in the upper equation of \eqref{eq:lms}. Hence, this variant is also called the normalized LMS (NLMS) filter. The LMS filter produces an output signal that is the best least-squares estimate of $n(k)$ in $x(k)$. It uses gradient descent to adjust the filter weights. 

\subsection{Pre-Whitening}
\label{sec:inverse}

The adaptive LMS filter operates best in presence of white noise. Any correlation between the elements in $\vec{n}_0(k)$ results in a slower convergence rate due to the associated correlation between the filter coefficients. The convergence rate for non-white, \ie, colored noise improves substantially if the data in $\vec{n}_0(k)$ is decorrelated. Thus, on the assumption that $n_0(k)$ is an autoregressive process of order $P$, we may resort to linear prediction to find a corresponding inverse filter, which can be then used to flatten the noise spectrum. This is accomplished as follows. The pre-whitened reference noise signal is 
\begin{equation} 
\tilde{n}_0(k) = n_0(k) - \hat{n}_0(k) = \vec{v}^\T \, \vec{n}_0(k) \text{,} 
\end{equation} 
where $\hat{n}_0(k)$ is a weighted sum of the past values of $n_0(k)$, 
\begin{equation} 
\hat{n}_0(k) = \sum_{p = 1}^P{a_p \, n_0(k - p)} \text{,} 
\label{eq:predictor} 
\end{equation} 
and $a_p$, $p = 1, 2, \dots, P$, are the predictor's coefficients. The inverse filter, accordingly,  has the following form 
\begin{equation}
\vec{v} = {\begin{bmatrix} 1 & -a_1 & -a_2 & \cdots & -a_P \end{bmatrix}}^\T \text{.} 
\end{equation}
For the sake of consistency, we proceed in the same way with the noise residual in the error signal using the same weights: 
\begin{equation} 
\tilde{e}(k) = e(k) - \sum_{p = 1}^P{a_p \, e(k - p)} = \vec{v}^\T \, \vec{e}(k) \text{.} 
\end{equation} 
Finally, the update rule for the LMS filter in \eqref{eq:lms} becomes 
\begin{equation} 
\vec{w}(k + 1) = \vec{w}(k) + \mu \, \tilde{\vec{n}}_0(k) \, \tilde{e}(k) \text{.} 
\end{equation} 
It should be noted that the error signal $e(k)$ in \eqref{eq:lms} is left as it, \ie\ untouched. 

What we achieve by decorrelating the elements in $\vec{n}_0(k)$ is a more circular mean-square-error (MSE) function, which, as an immediate consequence, speeds up the convergence. It is equivalent to normalizing and rotating the hyperboloid in a way that its principal axes align with the (orthogonal) axes of the parameter space $\R^M$ of $\vec{w}(k)$. Orthogonalization, or pre- whitening, also helps with non-stationary noise. In that case, however, the inverse filter must be tracked over time. Further details on pre-whitening and the LMS filter's efficiency with non-stationary inputs can be found in \cite{Vaseghi2008, Widrow1984_IT}. 

\subsection{Moving-Average Wiener Filtering}
\label{sec:maw}

The corresponding optimal filter is calculated as shown. It represents the solution to which the adaptive filter converges after a sufficiently large number of iterations under stationary conditions. Since the data sequence may be of infinite length, we calculate a different filter for each new block of data. 

Let $M$ successive samples of a reference noise, $n_0(k)$, be stored as a vector
\begin{equation}
\vec{n}_0(k) = {\begin{bmatrix} n_0(k) & n_0\rbracks{k - 1} & \cdots & n_0\rbracks{k - M + 1} \end{bmatrix}}^\T 
\end{equation}
with $\vec{n}_0(k) \in \R^M$. Convolving $n_0(k)$ with   
\begin{equation}
\vec{w}(k) = {\begin{bmatrix} w_0(k) & w_1(k) & \cdots & w_{M - 1}(k) \end{bmatrix}}^\T \text{,} 
\end{equation}
where $\vec{w}(k) \in \R^M$ is a transversal finite impulse response or FIR filter of order $M - 1$, we obtain 
\begin{equation}
y(k) = \vec{w}^\T(k) \, \vec{n}_0(k) \text{.} 
\label{eq:estimate}
\end{equation}
Now, extending this principle to a block of size $N$, we have 
\begin{equation}
\vec{y}(k) = {\begin{bmatrix} y(k) & y\rbracks{k + 1} & \cdots & y\rbracks{k + N - 1} \end{bmatrix}}
\end{equation}
with $\vec{y}(k) \in \R^{1 \times N}$, which is obtained according to 
\begin{equation}
\vec{y}(k) = \vec{w}^\T(k) \, \vec{N}_0(k) \text{,} 
\end{equation}
where $\vec{N}_0(k) \in \R^{M \times N}$, $M < N$, is a Toeplitz matrix, \ie\ 
\begin{equation}
\begin{aligned}
&\vec{N}_0(k) = \\ 
&\qquad {\begin{bmatrix} \vec{n}_0(k) & \vec{n}_0\rbracks{k + 1} & \cdots & \vec{n}_0\rbracks{k + N - 1} \end{bmatrix}} \text{.} 
\end{aligned} 
\end{equation}
The estimation error, or the output, is the difference
\begin{equation}
e(k) = x(k) - y(k) \text{,} 
\label{eq:error}
\end{equation}
and respectively 
\begin{equation}
\vec{e}(k) = {\begin{bmatrix} e(k) & e\rbracks{k + 1} & \cdots & e\rbracks{k + N - 1} \end{bmatrix}}
\end{equation}
with $\vec{e}(k) \in \R^{1 \times N}$. Equally, 
\begin{equation}
\vec{x}(k) = {\begin{bmatrix} x(k) & x\rbracks{k + 1} & \cdots & x\rbracks{k + N - 1} \end{bmatrix}}
\end{equation}
with $\vec{x}(k) \in \R^{1 \times N}$. Given \eqref{eq:estimate} and \eqref{eq:error}, we see that the signal $\hat{s}(k)$ in an arbitrary data block is given by the sequence 
\begin{equation}
\vec{e}(k) = \vec{x}(k) - \vec{w}^\T(k) \, \vec{N}_0(k) \equiv \hat{\vec{s}}(k) \text{.} 
\end{equation}

The LMS filter coefficients are adapted via minimization of $e^2(k)$, which corresponds to the minimization of the mean error power when $s(k)$, $n(k)$, and $n_0(k)$ are stationary. This on the other hand is equivalent to minimizing the mean noise power by matching the correlated noise $n_0(k)$ to $n(k)$ which is accomplished through $\vec{w}(k)$. The optimal weights are thus given by the Wiener--Hopf solution
\begin{equation}
\vec{w}_\textopt(k) = \vec{R}_{\vec{n}_0 \vec{n}_0}^{-1}(k) \, \vec{p}_{\vec{n}_0 x}(k) \text{,} 
\label{eq:optimum_filter}
\end{equation}
where $\vec{R}_{\vec{n}_0 \vec{n}_0}(k)$ is an auto-covariance matrix and $\vec{p}_{\vec{n}_0 x}(k)$ is a cross-covariance vector between $\vec{n}_0(k)$ and $x(k)$ \cite{Haykin2013}.

 In order to compute \eqref{eq:optimum_filter}, one would typically replace the variables $\vec{R}_{\vec{n}_0 \vec{n}_0}(k)$ and $\vec{p}_{\vec{n}_0 x}(k)$ by their sample estimates 
\begin{equation}
\widehat{\vec{R}}_{\vec{N}_0 \vec{N}_0}(k) = \frac{1}{N} \, \vec{N}_0(k) \, \vec{N}_0^\T(k)
\end{equation}
and 
\begin{equation}
\hat{\vec{p}}_{\vec{N}_0 \vec{x}}(k) = \frac{1}{N} \, \vec{N}_0(k) \, \vec{x}^\T(k) \text{.}
\end{equation}
This means that the filter $\vec{w}_\textopt(k)$ is computed over a window of the size $M + N - 1$. As the signal model presumes weak stationarity and ergodicity for $n(k)$, the hop size $L$ for $k$ can be set equal to $N$ to minimize computational cost. If $n(k)$ is non-stationary, however, the hop size, and so the size of  $\vec{x}(k)$ and $\vec{y}(k)$, should be reduced to a number that corresponds to the size of a segment over which $n(k)$ is stationary. In audio, $L$ can also be chosen according to the temporal resolution of the human ear. Note that it may appear that it is necessary to compute the filter itself using $M + N - 1$ samples, whereas the estimated sequence, $\hat{\vec{s}}(k)$, might be $L$ samples long. That would be the case, \eg, if $M \gg L$, and thus $N \gg L$ to make sure that $N > M$ for computational reasons. In the extremest case, $L = 1$, \ie\ the filtering is carried out sample-wise. 

From the considerations above, it should become evident that this technique is very expensive for non-stationary noise and for a high-order filter. And even though algorithms exist that do efficiently solve \eqref{eq:optimum_filter} via the Cholesky decomposition \eg, they still possess a considerable computational load and also require a great amount of memory. As a general rule, the order of the filter scales with the spectral dynamics of $n(k)$. 

\subsection{Spectral Subtraction}
\label{sec:subtract}

One of the issues related to the above technique is that when $s(k)$ and $n(k)$ are (locally) correlated, $y(k)$ will be estimated with a higher amplitude than the actual noise. This may lead to audible artifacts after subtracting $y(k)$ from $x(k)$. Another issue are abrupt changes, or jumps, of filter weights between consecutive blocks, which may cause clicks and pops. Linear interpolation is one possible approach to handle this. A more effective way to tackle both issues simultaneously is to carry out the subtraction in the the spectral domain according to 
\begin{equation}
\begin{aligned}
&\abs{E(\omega)} = \\
	&\qquad 
	\begin{dcases} 
		\sqrt[p]{\abs{X(\omega)}^p - \abs{Y(\omega)}^p} & \text{if}\ \abs{X(\omega)} > \abs{Y(\omega)} \text{,} \\
		0 & \text{otherwise.}
	\end{dcases} \\
&\arg{E(\omega)} = \arg{X(\omega)} \text{,} 
\end{aligned}
\label{eq:spectral_sub}
\end{equation}
where $\omega$ is the frequency, $\abs{\cdot(\omega)}$ refers to the magnitude, and $\arg{\cdot(\omega)}$ refers to the argument or phase at $\omega$. The spectra are computed using the short-time Fourier transform, \eg, and $p$ is typically set to 2. The time-domain signal $e(k)$ is obtained by applying the inverse of the transform to \eqref{eq:spectral_sub}. 

\section{Accompaniment Cancellation}
\label{sec:solution}

And now, based on the preceding considerations, we draw a parallel between adaptive noise cancellation and the related accompaniment cancellation problem. Although it appears to be technically similar to the echo cancellation problem, there is a difference. Put simply, echo is a natural phenomenon and echo cancellation is a two-way communication problem. Our case is a one-way communication problem, which serves the purpose of simultaneous self-assessment during practice and also enables a machine analysis of the solo performance.

\subsection{Extended Signal Model}

The previous signal model is adapted and further amended in order to comply with the use case. The recorded signal is
\begin{equation}
x(k) = h(k) \ast \sbracks{d(k) + s(k)} \text{,}
\label{eq:extended_model}
\end{equation}
where $d(k)$ is the desired solo, $s(k)$ the accompaniment, and $h(k)$ the impulse response of the microphone. The asterisk $\ast$ denotes convolution. More precisely, we consider $s(k)$ to be approximately equal to
\begin{equation}
s(k) \approx A \, g(k) \ast s_0 \rbracks{k - \kappa} \text{,}
\label{eq:accompaniment}
\end{equation}
\ie\ the result of the colorization of the reference $s_0(k)$ by the speaker $g(k)$, which is attenuated and delayed by $\kappa$ samples on its path to the microphone through the acoustic channel.

\begin{figure}[!ht]
\centering
\includegraphics[width=\columnwidth]{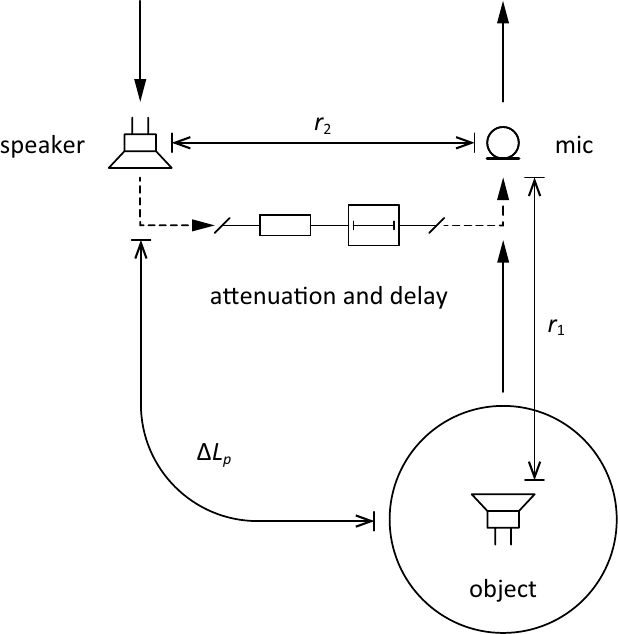}
\caption{Accompaniment cancellation problem.}
\label{fig:model}
\end{figure}

Modifications of the solo that are induced by the channel are ignored or are considered as being part of the signal. Any kind of additive white noise is omitted due to short distances between the sound sources and the microphone. We measure a signal-to-noise ratio (SNR) near 50 dB for the solo, \eg\ As before, we postulate statistical independence for the solo and the accompaniment. In addition, the propagation between the speaker and the microphone first and foremost takes place in and over the direct path, see  \eqref{eq:accompaniment} and Fig.~\ref{fig:model}.

\subsection{Convergence Analysis}

Applying the optimum solution from \eqref{eq:optimum_filter} to \eqref{eq:extended_model}, we have
\begin{equation}
\vec{w}_\textopt(k) = \vec{R}_{\vec{s}_0 \vec{s}_0}^{-1}(k) \, \vec{p}_{\vec{s}_0 x}(k) \text{.}
\label{eq:adapted_filter}
\end{equation}
Now, the sample cross-covariance vector writes
\begin{align}
\hat{\vec{p}}_{\vec{s}_0 x}(k) &= \frac{1}{N} \, \vec{S}_0(k) \, \vec{x}^\T(k) \\
	&= \frac{1}{N} \, \vec{S}_0(k) \, \sbracks{\vec{D}^\T(k) + \vec{S}^\T(k)} \, \vec{h} \nonumber \text{,}
\end{align}
and so
\begin{equation}
\vec{p}_{\vec{s}_0 x}(k) =  \sbracks{\vec{R}_{\vec{s}_0 \vec{d}}(k) + \vec{R}_{\vec{s}_0 \vec{s}}(k)} \, \vec{h} \text{,}
\label{eq:cross-covariance}
\end{equation}
where
\begin{equation}
\vec{h} = {\begin{bmatrix} h_0 & h_1 & \cdots & h_{M - 1} \end{bmatrix}} \in \R^M
\end{equation}
is the microphone's finite impulse response of order $M - 1$. The two Toeplitz matrices $\vec{D}(k)$ and $\vec{S}(k)$ are constructed in line with Fig.~\ref{fig:toeplitz}.
\begin{figure*}[!ht]
\begin{equation}
\vec{A}(k) = \begin{bmatrix}
	a(k) & a\rbracks{k - 1} & \cdots & a\rbracks{k - M + 1} & \cdots & a\rbracks{k - N + 1} \\
	\vdots & a(k) & \cdots & a\rbracks{k - M + 2} & \cdots & a\rbracks{k - N + 2} \\
	\vdots & \vdots & \ddots & \vdots & \vdots & \vdots \\
	0 & \vdots & \ddots & \vdots & \vdots & \vdots \\
	0 & 0 & \cdots & a(k) & \cdots & a\rbracks{k - N + M}
\end{bmatrix}
\in \R^{M \times N} \text{, } M < N
\end{equation}
\caption{Structure of a Toeplitz matrix used to formulate convolution as a matrix multiplication.}
\label{fig:toeplitz}
\end{figure*}
Using \eqref{eq:cross-covariance}, \eqref{eq:adapted_filter} becomes
\begin{equation}
\vec{w}_\textopt(k) = \vec{R}_{\vec{s}_0 \vec{s}_0}^{-1}(k) \, \sbracks{\vec{R}_{\vec{s}_0 \vec{d}}(k) + \vec{R}_{\vec{s}_0 \vec{s}}(k)} \, \vec{h} \text{.}
\label{eq:analytic_filter}
\end{equation}
From \eqref{eq:analytic_filter} it can be seen that if $\E\cbracks{s_0(k) \, d(k)} = 0 \, \forall \, k$ and if $\E\cbracks{s_0(k) \, s(k)} = \E\cbracks{s_0(k) \, s_0(k)} \, \forall \, k$, \ie\ if the channel has no influence on the accompaniment, the optimum filter $\vec{w}_\textopt$ is equal (converges) to the microphone's impulse response $\vec{h}$. It also means that the filter should be as long as $\vec{h}$. However, in practice, the filter must be computed using a finite sample, as indicated by the time index $k$ and the sample size $N$. And so, depending on the sample covariance between $d(k)$, $s(k)$, and $s_0(k)$, the filter may locally vary. Thus, the hop size between two samples must be kept sufficiently small to avoid artifacts at transition points. It should also be taken into consideration that the sample size $N$ should be about ten times longer than the filter length for it to converge towards the optimum. This, inevitably, comes along with a high computational load.
 
The solo estimate is equal to the estimation error, \ie
\begin{align}
e(k) &= x(k) - \vec{w}_\textopt^\T(k) \, \vec{s}_0(k) \\
	&\overset{\eqref{eq:analytic_filter}}{=} \vec{h}^\T \, \sbracks{\vec{d}(k) + \vec{s}(k)} \nonumber \\
	&\qquad - \vec{h}^\T \, \sbracks{\vec{R}_{\vec{s}_0 \vec{d}}(k) + \vec{R}_{\vec{s}_0 \vec{s}}(k)} \, \vec{R}_{\vec{s}_0 \vec{s}_0}^{-1}(k) \, \vec{s}_0(k) \nonumber \\
	&= \vec{h}^\T \, \underbrace{\sbracks{\vec{d}(k) - \vec{R}_{\vec{s}_0 \vec{d}}(k) \, \vec{R}_{\vec{s}_0 \vec{s}_0}^{-1}(k) \, \vec{s}_0(k)}}_{= \, \vec{d}(k)\ \text{iff}\ \E\cbracks{s_0(k) \, d(k)} \, = \, 0} \nonumber \\
	&\qquad + \vec{h}^\T \, \underbrace{\sbracks{\vec{s}(k) - \vec{R}_{\vec{s}_0 \vec{s}}(k) \, \vec{R}_{\vec{s}_0 \vec{s}_0}^{-1}(k) \, \vec{s}_0(k)}}_{= \, \vec{0}\ \text{iff}\ \E\cbracks{s_0(k) \, s(k)} \, = \, \E\cbracks{s_0(k) \, s_0(k)}} \text{.}
\label{eq:analytic_solo}
\end{align}
Eq.~\eqref{eq:analytic_solo} can be interpreted as follows. If the guitar player is mute, \ie\ $d(k) = 0$, and there is no difference between $s_0(k)$ and $s(k)$, \ie\ the speaker and the room are negligible, $e(k)$ is zero and so is the estimate $\hat{d}(k)$. If the speaker and the room are not negligible, the error depends on how strongly $s(k)$ is correlated with $s_0(k)$. The stronger, the smaller the error. If, however, the player is performing, the error is further subject to the correlation between the solo $d(k)$ and $s_0(k)$. Since the cross-correlation between independent sources is never zero for finite samples, we can expect the solo estimate $\hat{d}(k)$ to be additionally degraded to the extent of the correlation between the solo and the reference accompaniment. Either way, it can be seen that the signature of the microphone (colorization) is part of the estimate, \ie\ it is not equalized.

\subsection{Delay Compensation}

Delay compensation is an important issue because it helps ameliorate the result of the cancellation via maximization of the empirical cross-covariance between $s(k)$ and $s_0(k)$. Delay compensation can be done manually or also automatically in a preceding calibration procedure. One option is to play back and record the accompaniment without the solo and to check where the cross-correlation function attains its maximum. In the case at hand, where a tablet computer is utilized for both the playback and the recording, the microphone's distance is about 25 cm from the speaker. Given that the speed of sound in dry air at 20 ${}^\circ$C is around 343 m/s, the time delay amounts to 0.72 ms or 32 samples at a sampling rate of 44.1 kHz. But as the delay in this case is much smaller than the sample size, which counts several thousands of observations, it also can be ignored. A much greater time offset between $x(k)$  and $s_0(k)$ is due to the hardware---and generally it cannot be neglected. The exact latency of the speaker and the microphone usually can be found in the system preferences.

\subsection{Short-Time Subband Wiener Filtering}
\label{sec:sbw}

As an alternative to the standard technique from Section \ref{sec:maw}, we present a different technique for computing \eqref{eq:adapted_filter}. It is not only much faster, but also requires much less memory. More importantly, the  technique is real-time capable \cite{Gorlow2013_ASL}. 

We resort to the short-time Fourier transform (STFT) and compute the local spectra $S_0(\omega)$ and $X(\omega)$. Then we form $Z$ subbands on the equivalent rectangular bandwidth rate scale, which is given by \cite{Glasberg1990_Hearing}
\begin{equation}
\ERBS(f) = 21.4 \cdot \log_{10} \rbracks{1 + 4.37 \cdot f} \text{,} 
\end{equation}
where $f$ is the frequency in kHz. In the continuous case, $\omega$ is equivalent to $f$. The spectral components in the $\zeta$th subband can be represented by the corresponding centroids $S_0(\zeta)$ and $X(\zeta)$, $\zeta = 1, 2, \dots, Z$. This would be equivalent to making a sequence of $N$ samples pass through a non-uniform complex filter bank consisting of $Z$ auditory filters. The big advantage of the STFT is the availability of many optimized libraries to compute the underlying fast Fourier transform (FFT).\footnote{See, \eg, \url{http://www.fftw.org/}.}

The computation of the filter from \eqref{eq:adapted_filter} is as follows. The auto-covariance of $S_0(\omega)$ in subband $\zeta$ is 
\begin{equation}
R_{S_0 S_0}(\zeta) = \frac{1}{\abs{\Omega_\zeta}} \, \sum_{\inf{\Omega_\zeta}}^{\sup{\Omega_\zeta}}{\abs{S_0(\omega)}^2} \qquad \forall \, \omega \in \Omega_\zeta
\end{equation}
and the cross-covariance between $S_0(\omega)$ and $X(\omega)$ is 
\begin{equation}
P_{S_0 X}(\zeta) = \frac{1}{\abs{\Omega_\zeta}} \, \sum_{\inf{\Omega_\zeta}}^{\sup{\Omega_\zeta}}{\abs{S_0^\ast(\omega) \, X(\omega)}} \qquad \forall \, \omega \in \Omega_\zeta \text{,}
\label{eq:sub_x-cov}
\end{equation}
where superscript $\ast$ denotes complex conjugation. And so, 

\begin{equation}
W_\textopt(\zeta) = \frac{P_{S_0 X}(\zeta)}{R_{S_0 S_0}(\zeta)} \qquad \forall \, \zeta \text{.} 
\label{eq:wiener}
\end{equation}
Accordingly, the matched accompaniment is 
\begin{equation}
Y(\omega) = W_\textopt(\zeta) \, S_0(\omega) \qquad \forall \, \omega \in \Omega_\zeta \text{.} 
\label{eq:stft_estimate}
\end{equation}
The error $E(\omega)$ is calculated in accordance with \eqref{eq:spectral_sub}, where $p = 1$, and transformed back to time domain using either the overlap-add or the overlap-save method \cite{Oppenheim2009}. It should be evident that the proposed technique is much more efficient, because:
\begin{itemize}
	\item the block size of the STFT $N$ is relatively small,
	\item the filter order $Z$ (number of subbands) is low,\footnote{$Z$ is equivalent to $M$ from before.} 
	\item no matrix inversion is required, only division. 
\end{itemize}

%
%
%
%
%
%

\section{Diversity Combining}
\label{sec:diversity}

\begin{figure}[!ht]
\centering
\includegraphics[width=.6\columnwidth]{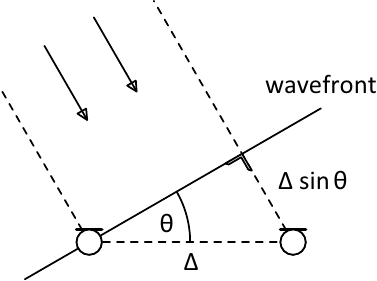}
\caption{Angle of arrival $\theta$, spacing $\Delta$, and spatial shift $\Delta \, \sin \theta$.}
\label{fig:diversity}
\end{figure}

Now if we depart from the condition of being restricted to the use of the built-in mic of a tablet computer, we may also consider other forms of external hardware. One alternative is a microphone array. So, let us focus on the case where there are two identical microphones. They shall be placed in close proximity. According to the Nyquist--Shannon  theorem, if we were to apply spatial filtering, their distance must be shorter than or equal to half the wavelength of the signal component with the highest frequency $f_{\max}$. In regard to our test data, a frequency of 8 kHz would be the upper limit.\footnote{The maximum frequency audible to the human ear is said to be 20 kHz. In many perception-related algorithms, however, it is reduced to 16 kHz.} And thus, the corresponding wavelength $\mathit{\lambda}_{\min}$ and distance $\Delta$ are
\begin{equation}
\lambda_{\min} = \frac{c}{f_{\max}} \leadsto \Delta = \frac{\lambda_{\min}}{2} = \frac{c}{2 \, f_{\max}} \text{,}
\end{equation}
where $c$ is the speed of sound. Accordingly, the spacing $\Delta$ is 2.1 cm for an $f_{\max}$ of 8 kHz. So, if we place the second mic along the line between the first mic and the solo instrument, the delay time between the two microphones amounts to $\nicefrac{1}{16}$ of a millisecond, or 3 periods at 44.1-kHz sampling. Also, if we presume that the instrument is 1 m (or more) away from the array, the impinging solo, in good approximation, has the same amplitude at both microphones. The resulting geometry is depicted in Fig.~\ref{fig:diversity}.

\subsection{Single-Input-Double-Output Signal Model}

Based on the above reflections, we amend \eqref{eq:extended_model} as follows:
\begin{equation}
{\begin{bmatrix} x_1(k) \\ x_2(k) \end{bmatrix}} = h(k) \ast {\begin{bmatrix} d_1(k) + s_1(k) \\ d_1(k - \kappa) + s_2(k)
\end{bmatrix}} \text{,}
\label{eq:sido}
\end{equation}
where $x_1(k)$ and $x_2(k)$ are the signals captured by the array. Only the delay between the two versions of the solo $d(k)$ is taken into account, as it is the signal we seek to enhance. In the Fourier domain, due to the shift theorem, \eqref{eq:sido} becomes
\begin{equation}
{\begin{bmatrix} X_1(\omega) \\ X_2(\omega) \end{bmatrix}} = H(\omega) \cdot {\begin{bmatrix} D_1(\omega) + S_1(\omega) \\ W_N^{\omega \kappa} D_1(\omega) + S_2(\omega) \end{bmatrix}}
\label{eq:sido_fft}
\end{equation}
with $W_N = \e^{-\im \nicefrac{2 \pi}{N}}$, where $\e^\cdot$ is the exponential function and $\im$ is the imaginary unit. The time shift is equivalent to a phase shift in the Fourier domain, which is a function of $\omega$.

\subsection{Delay Estimation}

Looking at \eqref{eq:sido_fft}, one can see that if only the instrument is active, \ie\ there is no accompaniment, the two output signals exhibit the following relation in the Fourier domain:
\begin{equation}
X_2(\omega) = W_N^{\omega \kappa} X_1(\omega) \text{.}
\label{eq:relation}
\end{equation}
In practice, when using the discrete Fourier transform, which is cyclic, \eqref{eq:relation} still holds largely true. It is because the delay $\kappa$ is much smaller than the transform size, which is 2048 or 4096 points in the general case. And so, the delay $\kappa$ may be estimated by taking the median of the below observations:
\begin{equation}
\hat{\kappa} = \underset{\omega}{\operatorname{median}} {\sbracks{-\arg \frac{X_2(\omega)}{X_1(\omega)} \, \frac{N}{2 \pi \, \omega}}} \qquad \forall \, \omega \text{.}
\end{equation}
Alternatively, the delay may be found by selecting the value with the highest number of occurrences in the corresponding histogram. Fig.~\ref{fig:diversity} also shows that the  delay can be associated with an angle of arrival $\theta$. The relation is
\begin{equation}
\frac{\kappa}{f_s} = \frac{\Delta \, \sin \theta}{c} \leadsto \theta = \arcsin {\rbracks{\frac{f_{\max}}{\nicefrac{f_s}{2}} \kappa}} \text{,}
\end{equation}
where $f_s$ is the sampling frequency and $f_{\max} \leqslant \nicefrac{f_s}{2}$.

\subsection{Maximal-Ratio Combining}

Maximal-ratio combining (MRC) is one known method of receiver diversity combining. It yields the optimum combiner for independent Gaussian interference. MRC is equivalent to the least-squares (LS) solution of the normal equations
\begin{equation}
{\rbracks{{\begin{bmatrix} 1 \\ W_N^{\omega \kappa} \end{bmatrix}}^\CT {\begin{bmatrix} 1 \\ W_N^{\omega \kappa}  \end{bmatrix}}}} \hat{D}(\omega) = {\begin{bmatrix} 1 \\ W_N^{\omega \kappa} \end{bmatrix}}^\CT {\begin{bmatrix} \hat{D}_1(\omega) \\ \hat{D}_2(\omega)  \end{bmatrix}} \text{,}
\end{equation}
where superscript $\CT$ denotes the Hermitian transpose and 
\begin{equation}
{\begin{bmatrix} \hat{D}_1(\omega) \\ \hat{D}_2(\omega) \end{bmatrix}} = {\begin{bmatrix} E_1(\omega) \\ E_2(\omega) \end{bmatrix}} \text{,}
\end{equation}
\ie, it is the output of cancelling the accompaniment in each channel independently according to Section \ref{sec:sbw}. The MRC solution can thus be formulated more explicitly as
\begin{equation}
\hat{D}(\omega) = \frac{1}{2} \, {\begin{bmatrix} 1 & W_N^{-\omega \kappa} \end{bmatrix}} {\begin{bmatrix} E_1(\omega) \\ E_2(\omega) \end{bmatrix}} \text{.}
\label{eq:mrc}
\end{equation}
Eq.~\eqref{eq:mrc} tells us that the signal from the second microphone is counter-rotated by the phase shift, so that the signals from both microphones are combined into a single estimate $\hat{D}(\omega)$, which yields the maximum ratio between the solo signal and the accompaniment residuals after subtraction \cite{Brennan1959}.

\section{Comparison}
\label{sec:simulation}

In this section, we simulate the accompaniment cancellation problem using prerecorded guitar solos in order to assess and compare the solutions that are elaborated in the previous two sections. We consider the speaker to be 25 cm away from the microphone and the sound object, \ie\ the guitar amplifier, to be in 100 cm distance. According to the distance law,
\begin{equation}
L_{p_2} - L_{p_1} = 20 \, \log_{10} \frac{r_1}{r_2} \text{,}
\end{equation}
we conclude that the sound pressure level (SPL) between the speaker and the guitar amp differs by $\Delta L_p \approx 12.0$ dB. Now, if we postulate that the SPL of the accompaniment is 6.02 dB below the SPL of the guitar amp, the root-mean-square value of the accompaniment in the recorded signal must be 6.02 dB higher than the level of the solo. Such a setting should allow the guitar player, who is deemed to be in the proximity of the amp, to hear his own performance. Note, however, that in the recorded mixture the accompaniment is twice as loud. Other phenomena, such as reverberation or noise, are neglected due to the short distances between the sound sources and the mic and also for the sake of simplicity. The remaining parameters in \eqref{eq:extended_model} and \eqref{eq:accompaniment} are chosen as follows: the channel delay $\kappa$ is equivalent to 0.72 ms, $g(k)$ is ignored, and $A$ is arbitrary but distinct from the accompaniment level in the mixture. Sony's C37-FET condenser microphone with a reverberation time of 13.7 ms or 606 samples at 44.1-kHz sampling is modeled by the impulse response $h(k)$. Respectively, we choose the filter length $M$ as the next larger power of 2 to leave a margin. As for the reference solo, it is obtained by convolving $h(k)$ with a prerecorded solo signal $d(k)$, see \eqref{eq:extended_model}. The simulations are run in MATLAB under Windows 8.1 on 20-s mixtures. With respect to the case where we simulate the use of an array of two microphones, their distance is 2.14 cm and the angles of arrival are $21.3^\circ$ and $90.0^\circ$, for the solo and accompaniment, respectively.

\subsection{Algorithms}

For comparison, we employ the following algorithms with the following settings. The chosen values were found to give subjectively the best result for each of the algorithms.
\begin{description}
\item[ANC] adaptive noise cancellation (see Section~\ref{sec:lms}),
	\begin{description}[labelwidth=.8cm]
	\item[$M$] 1023 (filter order)
	\item[$\mu$] 0.10 (step size)
	\end{description}
\item[ANC\textsuperscript{$\prime$}] ANC with inverse filtering (see Section~\ref{sec:inverse}),
	\begin{description}[labelwidth=.8cm]
	\item[$M$] 1023 (filter order)
	\item[$\mu$] 0.01 (step size)
	\item[$P$] 15 (inverse filter order)
	\end{description}
\item[MAW] moving-average Wiener filtering (see Section~\ref{sec:maw}),
	\begin{description}[labelwidth=.8cm]
	\item[$M$] 1023 (filter order)
	\item[$N$] 16384 (sample size)
	\item[$L$] 64 (hop size)
	\end{description}
\item[MAW\textsuperscript{$\prime$}] MAW with spectral subtraction (see Section~\ref{sec:subtract}),
	\begin{description}[labelwidth=.8cm]
	\item[$M$] 1023 (filter order)
	\item[$N$] 16384 (sample size)
	\item[$L$] 64 (hop size)
	\item[\textnormal{$N_\textrm{FFT}$}] 4096 (FFT size)
	\item[\textnormal{$L_\textrm{FFT}$}] 2048 (FFT hop size)
	\end{description}
\item[SBW] subband Wiener filtering (see Section~\ref{sec:sbw}), and
	\begin{description}[labelwidth=.8cm]
	\item[\textnormal{$N_\textrm{FFT}$}] 4096 (FFT size)
	\item[\textnormal{$L_\textrm{FFT}$}] 2048 (FFT hop size)
	\item[$Z$] 39 (number of subbands)
	\end{description}
\item[SBW\textsuperscript{$\prime$}] single-input-double-output SBW (see Section~\ref{sec:diversity}).
	\begin{description}[labelwidth=.8cm]
	\item[\textnormal{$N_\textrm{FFT}$}] 4096 (FFT size)
	\item[\textnormal{$L_\textrm{FFT}$}] 2048 (FFT hop size)
	\item[$Z$] 39 (number of subbands)
	\end{description}
\end{description}
\subsection{Metrics}
As far as an objective quality assessment is concerned, it can be said that there is no consensus across the domain about an ever applicable metric for audio enhancement algorithms. As for the various metrics that exist to assess speech quality, one should be aware that they apply only with restrictions. Music is much more complex than speech, to put it crudely. Metrics that came out from the source separation community are still far from the reality of perceived audio quality. Partially, they show a weak correlation or even contradict our perception as indicated in \cite{Gorlow2013_MLSP}. For these reasons, we resort to the following metrics. They have shown a certain degree of consistency on numerous occasions. These (and some other) metrics are:
\begin{description}[leftmargin=1.3cm,labelwidth=1.1cm]
\item[RMSD] the mean root-mean-square deviation averaged over non-overlapping data blocks of 23 ms duration,
	\begin{equation}
	\begin{aligned}
		\RMSD &= \frac{1}{T \sqrt{N}} \, \sum_{\tau = 1}^T {\left\{ \sum_{k = 0}^{N - 1} {\left[ \hat{d}(k, \tau) \right.} \right.} \\
			&\qquad {\left. {\left. {} - h(k) \ast d(k, \tau) \right]}^2 \right\}}^{\nicefrac{1}{2}} \text{,}
	\end{aligned}
	\end{equation}
	where $N$ is the block size and $T$ is their number,
\item[SNRF] a mean frequency-weighted signal-to-noise ratio \cite{Tribolet1978, Gorlow2013_ASL} averaged over frequency bands and segments,
	\begin{equation}
		\SNRF = \frac{10}{T \, Z} \, \sum_{\tau = 1}^T \sum_{\zeta = 1}^Z 
			\log_{10} \frac{\Psi_\mathrm{S}(\zeta, \tau)}{\Psi_\mathrm{N}(\zeta, \tau)} \text{,}
	\end{equation}
	where
	\begin{equation}
		\Psi_{\mathrm{S}}(\zeta, \tau) = \frac{1}{\abs{\Omega_\zeta}} \, 
			\sum_{\inf{\Omega_\zeta}}^{\sup{\Omega_\zeta}}
			\abs{H(\omega) \, D(\omega, \tau)}^2
		\label{eq:signal}
	\end{equation}
	$\forall \, \omega \in \Omega_\zeta$ and
	\begin{equation}
	\begin{aligned}
		\Psi_{\mathrm{N}}(\zeta, \tau) &= \frac{1}{\abs{\Omega_\zeta}} \, 
			\sum_{\inf{\Omega_\zeta}}^{\sup{\Omega_\zeta}}
			{\left\{ {\left[ \abs{\hat{D}(\omega, \tau)} \right.} \right.} \\
		&\qquad {\left. {\left. {} - \abs{H(\omega) \, D(\omega, \tau)} \right]}^2 \right\}}
	\end{aligned}
	\label{eq:noise}
	\end{equation}
	$\forall \, \omega \in \Omega_\zeta$, respectively,
\item[MOS] a mean opinion score computed with a basic version of PEAQ\footnote{Acronym for ``Perceptual Evaluation of Audio Quality''.} \cite{Kabal2003, PQevalAudio}, and finally
\item[RTF] the real-time factor, defined as the execution time of the algorithm divided by the signal's duration.\footnote{Measured on an Intel Core i7-4510U CPU operating at 2.00 GHz.}
\end{description}

\subsection{Results}

\begin{figure*}[!ht]
\includegraphics[width=\textwidth]{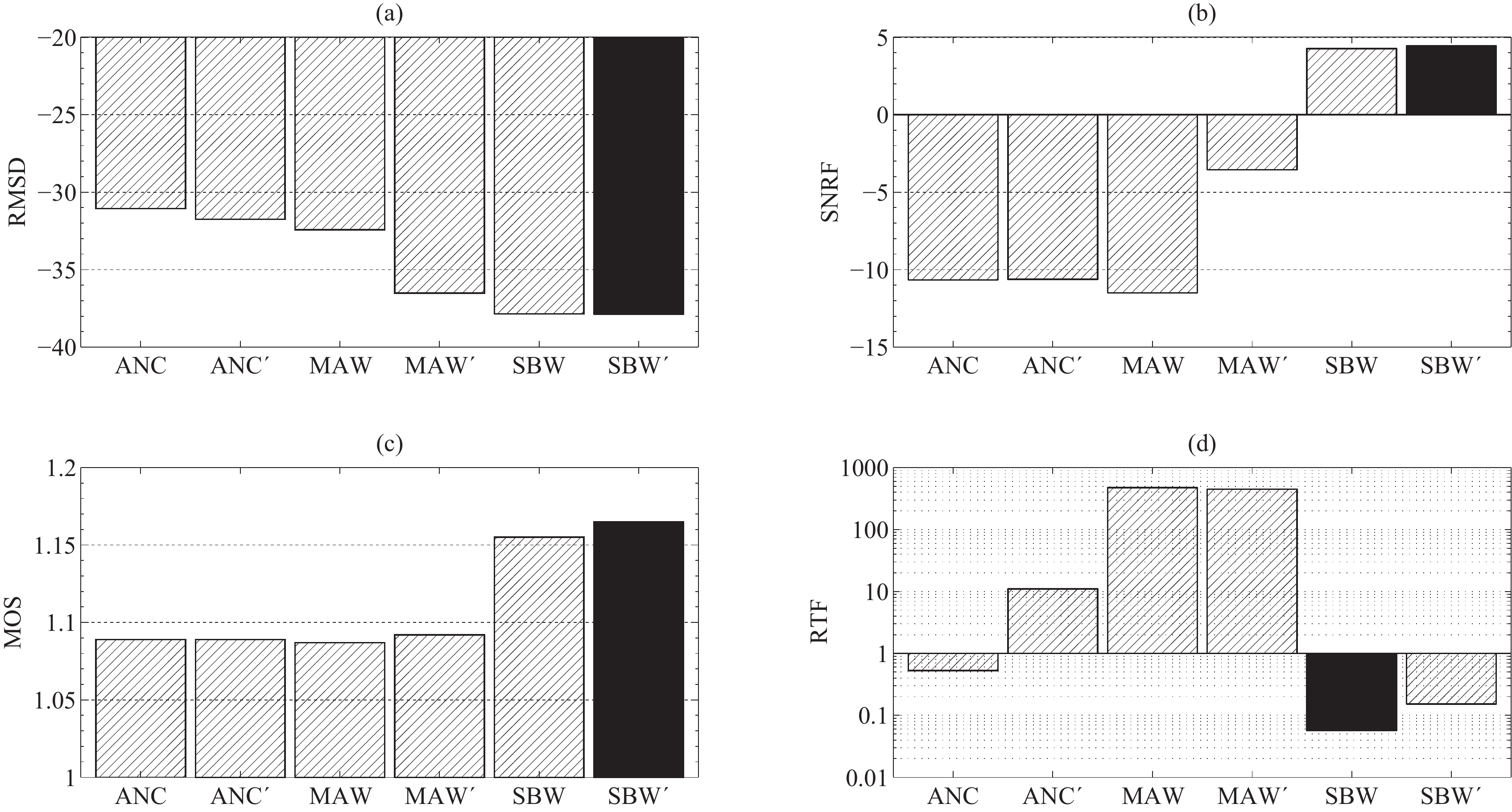}
\caption{Comparison between various filtering approaches \wrt\ the chosen performance metrics for ``All Blues''. Best values correspond to \textbf{black} bars.}
\label{fig:comparison}
\end{figure*}

The results for one mixture are listed in Table~\ref{tab:comparison}. It can be seen that with pre-whitening the numerical similarity slightly improves for adaptive filtering. This, however, has not such a significant impact on the perception-related SNRF metric and the objective MOS, which is the same. The RTF, on the other hand, increases by a factor of 20 and the algorithm no longer runs in real time. Although numerically closer to the original than adaptive filtering, moving-average Wiener filtering in its basic form has a lower perceptual quality. A significant jump both numerically and perceptually can be observed when the matched signal is subtracted from the mixture in the spectral domain. But yet, the improvement surely does not justify the awfully long run time, which is around 400--500 times longer than real time. By far, the clear winner here is the ERB-band Wiener filtering algorithm. It shows the best accompaniment suppression performance, the highest perceptual quality, and it also has a virtually negligible run time. Diversity improves the estimate only by a narrow margin at the cost of a double execution time (still faster than the adaptive noise canceller). It can be explained by the fact that for the greater part MRC corrects the phase of the estimate, which then again is much less critical as an error source than the magnitude, especially in terms of our perception. As a final remark, we would like to draw the reader's attention to the consistency between the SNRF and MOS trends over all algorithms. Fig.~\ref{fig:comparison} illustrates the results from Table~\ref{tab:comparison} in a bar diagram.

\begin{table}
\caption{Simulation results for ``All Blues''. Best values are printed in bold.}
\renewcommand{\arraystretch}{1.1}
\centering
\begin{tabular}{l*{6}{c}}
\toprule
& ANC & ANC\textsuperscript{$\prime$} & MAW & MAW\textsuperscript{$\prime$} & SBW & SBW\textsuperscript{$\prime$}\\
\midrule
RMSD [dB] & $-31.0$ & $-31.7$ & $-32.3$ & $-36.4$ & $-37.7$ & $\mathbf{-37.8}$\\
SNRF [dB] & $-10.6$ & $-10.5$ & $-11.4$ & $-3.50$ & $4.30$ & $\mathbf{4.40}$ \\
MOS & $1.089$ & $1.089$ & $1.087$ & $1.092$ & $1.155$ & $\mathbf{1.165}$ \\
RTF & $0.53$ & $11.0$ & $475$ & $449$ & $\mathbf{0.06}$ & $0.15$ \\
\bottomrule
\end{tabular}
\label{tab:comparison}
\end{table}

\section{Evaluation}

Here in this section, we carry out a thorough evaluation of the proposed approach that is based on short-time ERB-band Wiener filtering plus spectral subtraction. For this, and if not otherwise specified, we use the following parameters.
\begin{description}
\item[SISO] Single-input-single-output:
\begin{itemize}
\item 4096-point fast Fourier transform,
\item 2048-point overlap (50 \%),
\item 4096-point Kaiser-Bessel derived (KBD) window,
\item Standard window shape,
\item Classical Wiener filter,
\item Equivalent-rectangular-bandwidth (ERB) scale,
\item Manhattan distance (1-norm),
\item Equal RMS levels for solo and accompaniment,
\item 32-sample channel delay (uncompensated).
\end{itemize}
\item[SIDO] Single-input-double-output:
\begin{itemize}
\item 2.14-cm distance between microphones,
\item Directional alignment with the instrument,
\item Solo coming from the front (90\textsuperscript{$\circ$}) and accompaniment from the side (0\textsuperscript{$\circ$}).
\end{itemize}
\end{description}
The SIDO case is an extension of the SISO case, and so, all the SISO parameters are the same for both cases. The results of the evaluation are summarized in Figs.~\ref{fig:tf_resolution}--\ref{fig:spatial_align}. Only the first two metrics, the RMSD and the SNRF, are further used.

\subsection{Time and Frequency Resolution}

\begin{figure*}[!ht]
\subfiguretopcaptrue
\subfigure[Transform size]{\includegraphics[width=\textwidth]{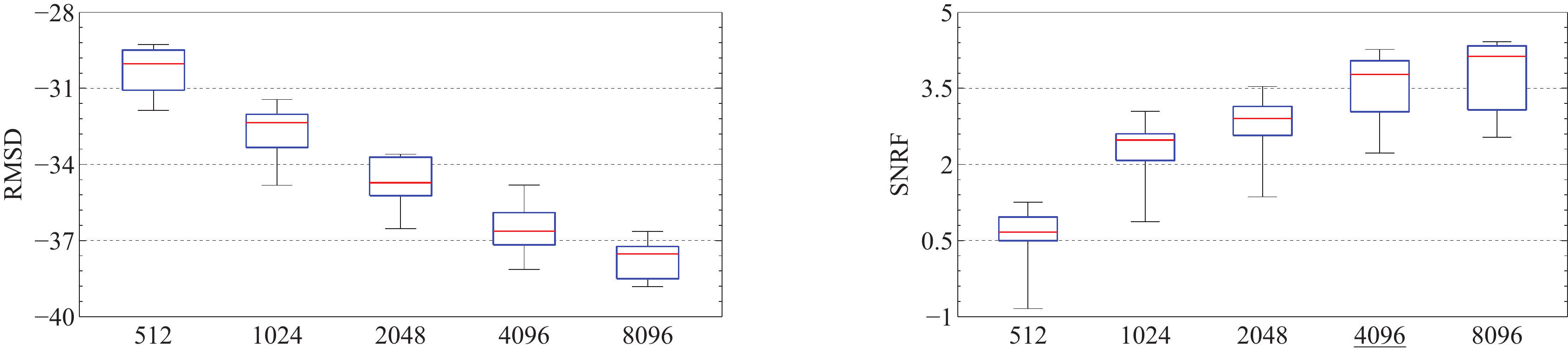}\label{fig:size}}
\subfigure[Window shape]{\includegraphics[width=\textwidth]{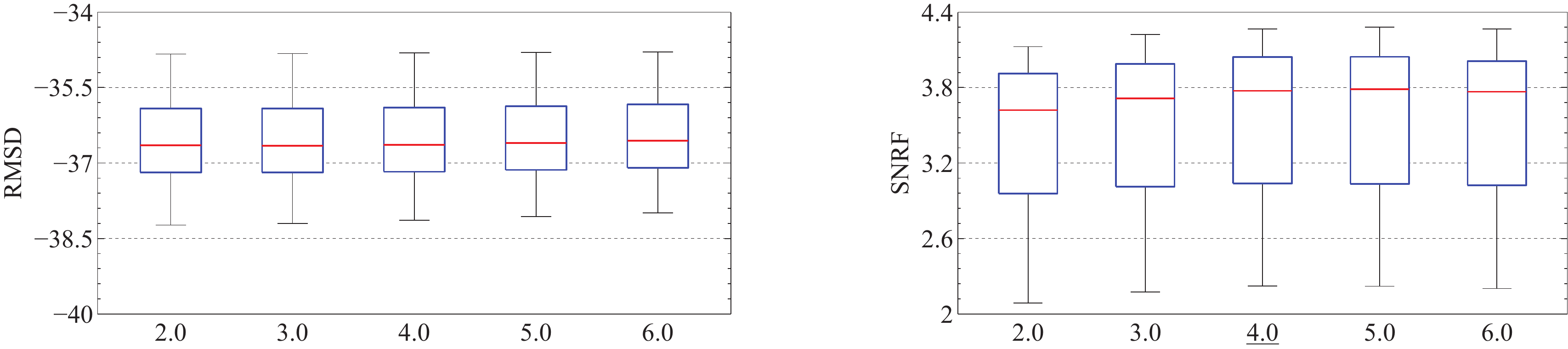}\label{fig:shape}}
\subfigure[Number of subbands]{\includegraphics[width=\textwidth]{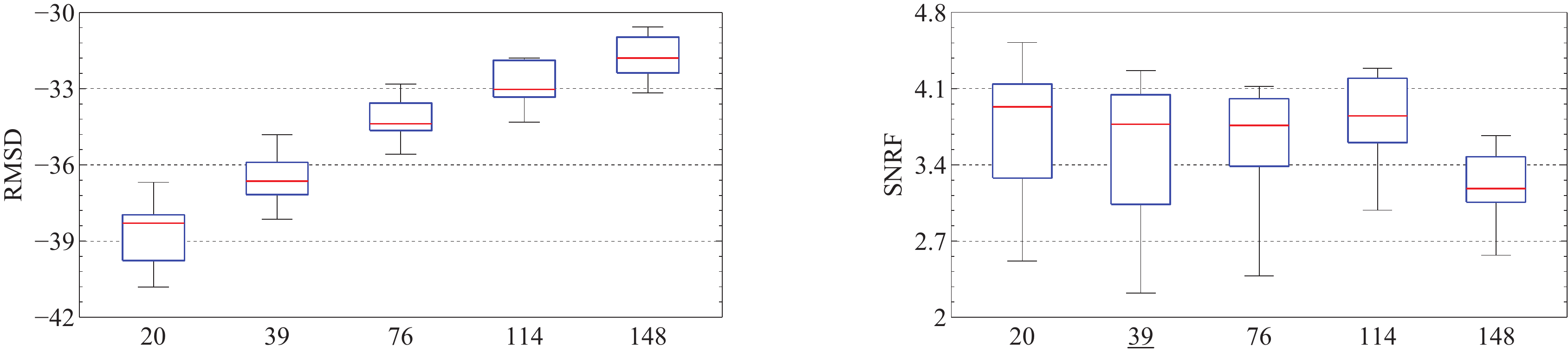}\label{fig:numsubbs}}
\caption{Algorithm performance as a function of (a) the FFT size, (b) the shape of the Kaiser-Bessel derived window \wrt\ its free parameter, and (c) the number of non-uniform subbands on a (pseudo-)ERB scale. Suggested values are \underline{underlined}. On each box, the central mark is the median and the edges of the box are the 25th and 75th percentiles. The whiskers extend to the most extreme data points including outliers.}
\label{fig:tf_resolution}
\end{figure*}

Fig.~\ref{fig:size} gives rise to the impression that the similarity of the estimate increases with the length of the transform. If we look at the SNRF alone, however, we notice that it begins to saturate after a value of 4096. If we listen to the estimate, it is clearly audible that the attacks of the played notes largely disappear if we further increase the window. For this and for the reason of memory requirements, we suggest a 4096-point DFT. The shape of the window appears to be a minor issue. Thus, we recommend to use the standard value for its shape parameter, which is 4, see also Fig.~\ref{fig:shape}.

With regard to the division of the frequency axis into non-uniform subbands, we may say the following: the number of subbands depends on one's preference and the application. A lower number preserves more of the original signal, but also lets more interference leak in. A higher number, on the other hand, deals better with interference, yet takes away the crisp details from the original sound. For this reason, a too low or too high number of subbands should be avoided. We suggest to use the standard  ERB scale that has 39 bands at 44.1-kHz sample rate and 16-kHz cutoff, see also Fig.~\ref{fig:numsubbs}.

\subsection{Filtering and Subtraction}

\begin{figure*}[!ht]
\subfiguretopcaptrue
\subfigure[Wiener filter exponent]{\includegraphics[width=\textwidth]{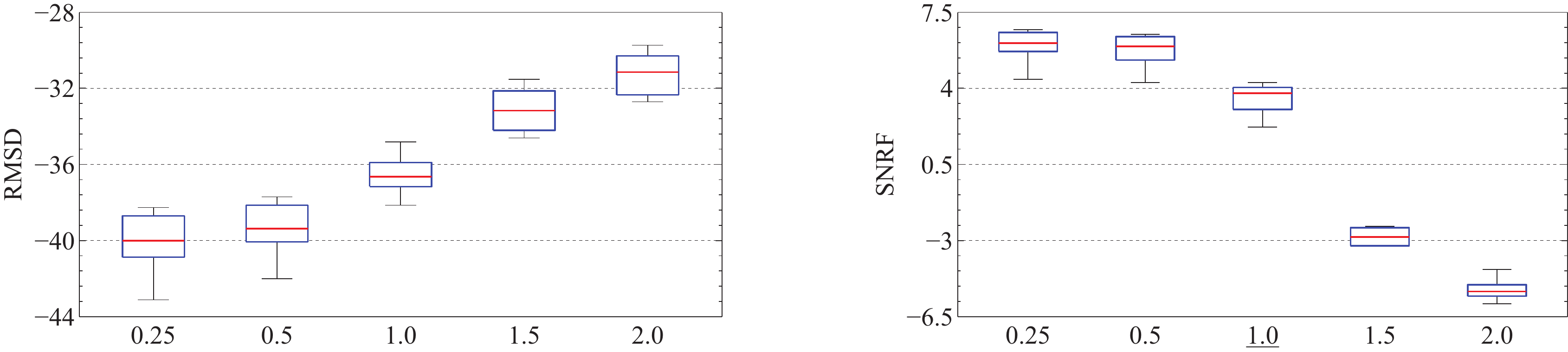}\label{fig:exponent}}
\subfigure[Spectral $p$-norm]{\includegraphics[width=\textwidth]{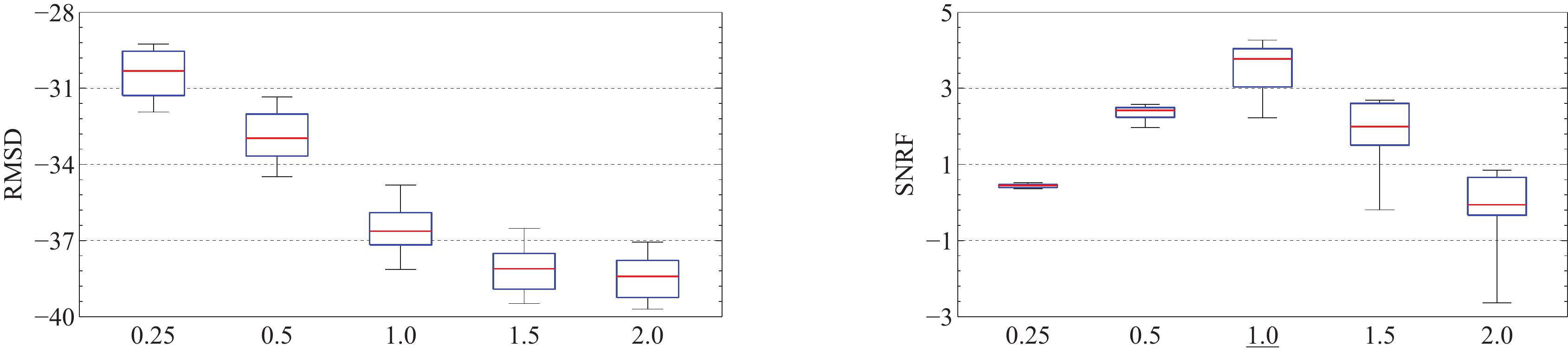}\label{fig:norm}}
\caption{Algorithm performance as a function of (a) the exponent of the Wiener filter and (b) the spectral $p$-norm of the estimate, see \eqref{eq:spectral_sub}. An exponent equal to $1.0$ yields the Wiener filter in \eqref{eq:wiener}. An exponent equal to $0.5$ yields the square-root Wiener filter. A $p$-value of 2 represents the Euclidean distance and a $p$-value of 1 represents the Manhattan distance, respectively. Suggested values are \underline{underlined}. On each box, the central mark is the median and the edges of the box are the 25th and 75th percentiles. The whiskers extend to the most extreme data points including outliers.}
\end{figure*}

What is said in the preceding paragraph also applies to the Wiener filter, \ie, there is a compromise to make. If the goal is to reduce the estimate's spectrum to its fewer components with a higher signal-to-interference ratio, the filter should be taken to a power lower than 1, see Fig.~\ref{fig:exponent}. However in our opinion, the classical Wiener filter is a safer bet, since it can sufficiently attenuate interference, while keeping most of the signal's fine structure intact.

On the contrary, Fig.~\ref{fig:norm} leaves no doubt about the most performant distance metric. It is the 1-norm, which is known as the Manhattan distance. Note that the 1-norm outperforms the 2-norm perceptually, even though the 2-norm is naturally associated with the underlying statistical model.

\subsection{Distance and Channel Delay}

\begin{figure*}[!ht]
\subfiguretopcaptrue
\subfigure[RMS level difference]{\includegraphics[width=\textwidth]{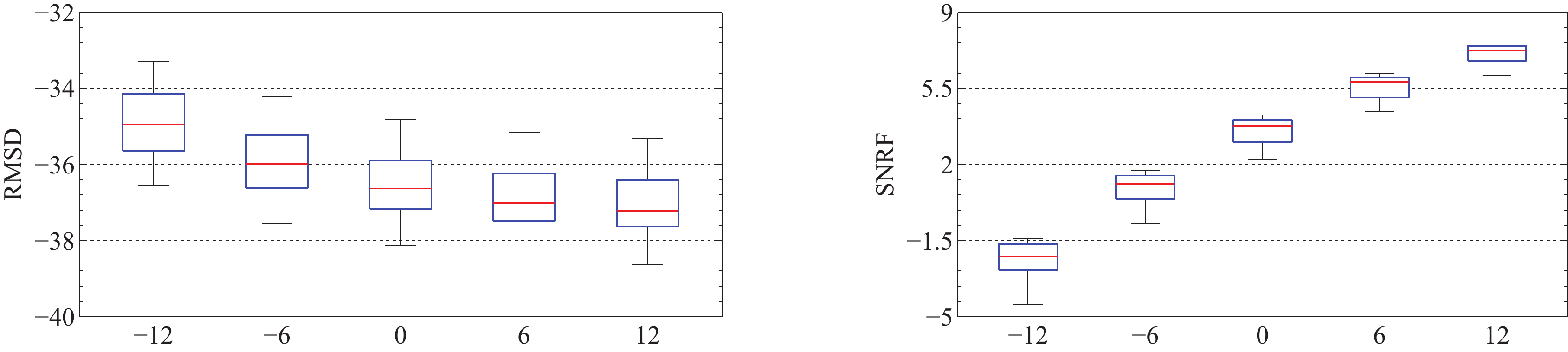}\label{fig:rms}}
\subfigure[Delay mismatch]{\includegraphics[width=\textwidth]{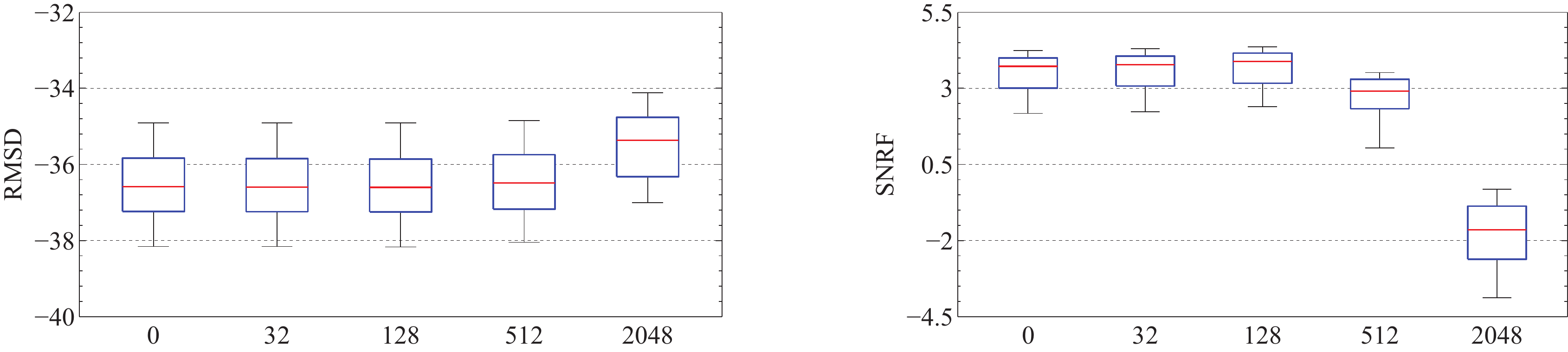}\label{fig:delay}}
\caption{Algorithm performance as a function of (a) the RMS level (or SPL) difference in decibel between the recorded solo and accompaniment and (b) the delay (samples) mismatch in the channel model. On each box, the central mark is the median and the edges of the box are the 25th and 75th percentiles. The whiskers extend to the most extreme data points including outliers.}
\end{figure*}

Fig.~\ref{fig:rms} simply confirms the blatantly obvious: the quality of the estimate is much higher, when the instrument is closer to the microphone or louder than the speaker. This leads to a higher solo-to-accompaniment ratio in the recording, and so, the subtracted accompaniment estimate leaves less traces.

More interesting, however, is Fig.~\ref{fig:delay}. It says that as long as the channel delay between the speaker and the microphone is much shorter than the transform length, its impact is null. This can be assumed to be the case for a tablet computer. It should still be noted that hardware delay may have a (much) greater impact depending on its latency time.

\subsection{Spatial Alignment}

\begin{figure*}[!ht]
\subfiguretopcaptrue
\subfigure[Distance between microphones]{\includegraphics[width=\textwidth]{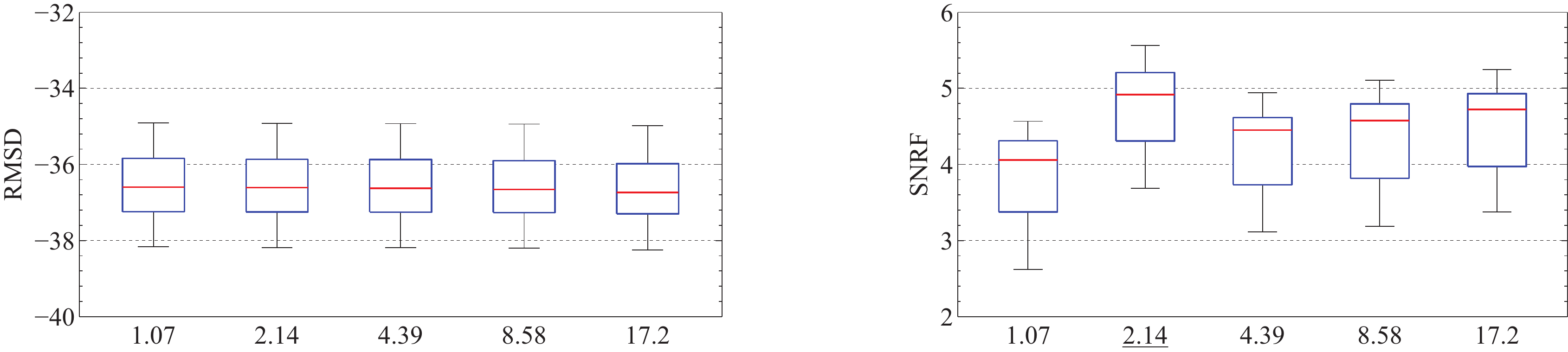}\label{fig:distance}}
\subfigure[Angular mismatch]{\includegraphics[width=\textwidth]{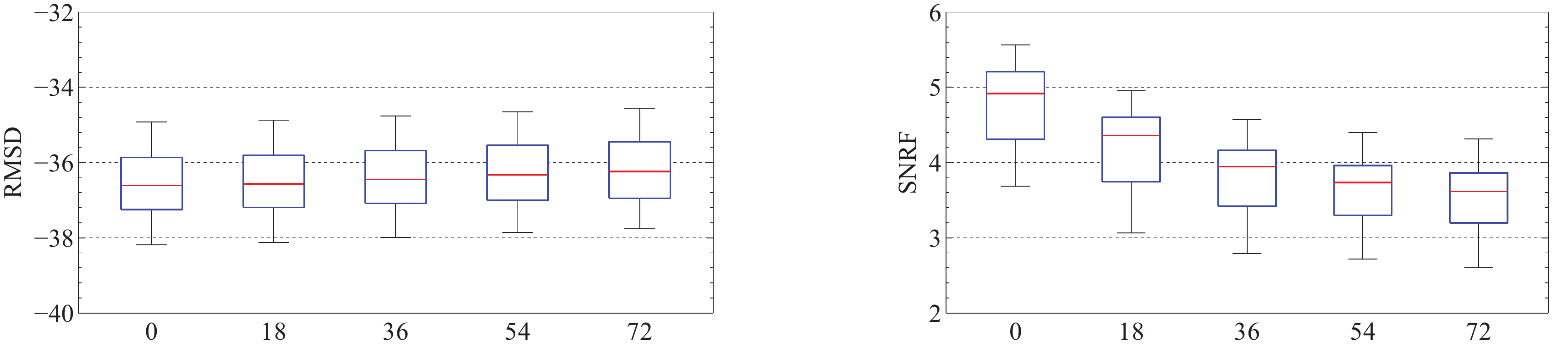}\label{fig:doa}}
\caption{Algorithm performance as a function of (a) the distance between the microphones and (b) the angular mismatch in degrees between the true and the estimated direction of arrival of the solo. The underlined value corresponds to half-wave spacing for a frequency of 8 kHz at 44.1-kHz sample rate. On each box, the central mark is the median and the edges of the box are the 25th and 75th percentiles. The whiskers extend to the most extreme data points including outliers.}
\label{fig:spatial_align}
\end{figure*}

Looking at Fig.~\ref{fig:distance} it seems that in the case where a pair of microphones is utilized to capture the composite signal in conjunction with maximal-ratio combining, the spacing is of vital importance. In our test, MRC performed best when the spacing between the microphones was equal to half the wave length of an 8-kHz signal. Especially \wrt\ transient artifacts, the enhancement was clearly noticeable.

Finally, Fig.~\ref{fig:doa} shows that a significant deviation in the angle estimate (related to the position of the guitar amplifier) is detrimental to sound quality. Nevertheless, a small error is certainly tolerable in most cases.

\subsection{General Remarks}

While the RMSD indicates a numerical deviation between the estimated and the desired signal, the SNRF is capable of capturing the perceptual component of an audio signal. If we look at the figures where we underlined the values for which we think the algorithm works best, we may observe that our perception correlates with the SNRF metric quite strongly. In consequence, we propose to employ the SNRF as a  standard metric for similar tasks. Nevertheless, we would like to point out that the SNRF has a tendency to give a high score to an estimate in which the noise or interference is suppressed to a very high degree. The corresponding signal may be perceived as condensed or less natural by the listener. From a machine perspective then again, the signal might be easier to analyze, since it contains the most dominant components, and what is more, the latter are associated with a low error.

In regard to the improvement that is achievable with a pair of microphones, or a microphone array in general, we would like to add the following. MRC as a post-processor is meant to correct the phase of the estimate after the accompaniment was subtracted from each channel separately. This has for effect a reduction of transient interference. As a rule, these transients are suppressed insufficiently by the core algorithm, because
\begin{itemize}
\item the transform is applied to a long window and
\item the sample covariance between the spectra of a note and a percussive hit has a high magnitude.
\end{itemize}
Altering the phase of the estimate, MRC essentially destroys the phase of the transient residual, which hence becomes not so or simply less prominent. For this to work, the array must be aligned with the spatial acoustics, as shown in Fig.~\ref{fig:spatial_align}. The achievable gain appears to be meager, particularly in view of the fact that additional hardware is necessary. The pragmatic solution would consist in using a guitar pickup or placing an external microphone close to the instrument or the amplifier, so that the recording exhibits a low interference. This would improve the estimate more than MRC, see Fig.~\ref{fig:rms}. A more sophisticated solution, which is yet to be found, would be to carry out the accompaniment cancellation during the optimal combination of the signals coming from the array.

\section{Conclusion}
\label{sec:conclusion}

Music signals pose a true challenge for the existing theory on statistical signal processing. This is due to its band width, its spectral dynamics, and its non-stationarity. Our paper shows that the convergence rate of the LMS filter is not fast enough to keep pace with transient music signals. Inverse filtering is a help but costs too much and the result is not  satisfactory. A Wiener-type filter is better, but only if the difference signal is computed in the frequency domain. It is, however, intractable from a computational point of view. A short-time Wiener can be viewed as the best and the cheapest solution, especially if the filtering is carried out in ERB-bands. Any other Bayesian estimator may also be employed, of course.

When using a microphone array, maximal-ratio combining can reduce the audible artifacts due to residual transients that reside in the solo after accompaniment cancellation. For this, the array should be aligned with the instrument. The spacing between the array elements is also important. The quality of the solo mainly depends on the distance of the sound source \wrt\ the sink: The closer the instrument to the mic or, more generally speaking, the louder the instrument in the recorded signal, the higher the quality of the solo. On a final note, the SNRF appears to be sufficiently correlated with perception to predict relative tendencies. Therefore, we recommend it as a reference metric for the assessment of quality for speech and audio.

\ifCLASSOPTIONcaptionsoff
  \newpage
\fi




\bibliographystyle{IEEEtran}
\bibliography{references}
%
%

%






\end{document}